\documentclass[aip,graphicx]{revtex4-1}

\usepackage{graphicx}
\usepackage{amssymb,amsfonts,amsmath,multirow}

\usepackage{color} 

\begin{document}


\title{An upper bound on community size in scalable community detection} 



\author{Gautier Krings}
\email[]{gautier.krings@uclouvain.be}
\affiliation{Institute ICTEAM, Universit\'e catholique de Louvain, B-1348 Louvain-la-Neuve, Belgium}

\author{Vincent D. Blondel}
\email[]{vincent.blondel@uclouvain.be, blondel@mit.edu}
\affiliation{Institute ICTEAM, Universit\'e catholique de Louvain, B-1348 Louvain-la-Neuve, Belgium}
\affiliation{Massachusetts Institute of Technology, Laboratory for Information and Decision Systems, 77 Massachusetts Avenue, Cambridge, MA 02139, USA}


\date{\today}

\begin{abstract}
It is well-known that community detection methods based on modularity optimization often fails to discover small communities. Several objective functions used for community detection therefore involve a resolution parameter that allows the detection of communities at different scales. We provide an explicit upper bound on the community size of communities resulting from the optimization of several of these functions. We also show with a simple example that the use of the resolution parameter may  artificially force the complete disaggregation of large and densely connected communities. 
\end{abstract}


\maketitle 

Many popular methods for detecting communities in networks are based on the optimization of the modularity function, which is a measure of the quality of a network partition into communities. The modularity of a partition compares the density of edges inside communities to the corresponding density expected in a null model \cite{newman_modul_PNAS}. It has been shown by Fortunato and Barth\'elemy \cite{F07} that modularity suffers from a so-called resolution limit: modularity optimization methods often fail to identify small communities.

Several authors have proposed objective functions for community detection that incorporate a tunable resolution parameter so as to allow community detection at different scales. One such function introduced by Reichardt and
Bornholdt in 2006  \cite{R06}, can be written in the following form:
\begin{equation} \label{modularity}
Q_\gamma =  \sum_{s} \frac{l_s}{L}-\gamma \left( \frac{d_s}{2L}\right)^2.
\end{equation}
In this expression, the sum is over the communities $s$, $l_s$ is the number of edges inside community $s$, $d_s$ is the sum of the degrees of the nodes in partition $s$ and $L$ is the total number of edges in the network. The case  $\gamma=1$  corresponds to modularity for the configuration model as defined by Newman \cite{newman_modul_PNAS}. Higher resolutions are obtained by choosing higher values for the resolution parameter $\gamma$ in Equation \ref{modularity}. Objective function  that are mathematically equivalent to $Q_\gamma$  have been proposed in a number of other contexts. In particular, Lambiotte et al. have shown \cite{L09} that the function $Q_\gamma$  corresponds to the first-order approximation of a dynamical process driven by the Laplacian of the graph where the resolution parameter plays the role of a timescale. The function $Q_\gamma$ is also a special case (for $\omega = 0$) of the function used by Mucha et al. \cite{M10} to study so-called multislice networks. It has been shown by Kumpala et al.  \cite{K07} that methods based on the optimization of $Q_\gamma$ suffer from a resolution limit similar to the one reported \cite{F07} for $\gamma =1$. 

In this note, we show that any resolution parameter value  $\gamma>1 $ impose a non-trivial upper bound on the size of  communities.  To establish this bound, consider  two communities whose node degrees sum to, respectively, $d_1$ and $d_2$ and contain, respectively, $l_1$ and $l_2$ internal edges. Let also $e$ be the number of edges connecting the two communities. Compare now the situation where the communities are separate with the one where the two communities are merged into one. In the latter case, the total degree of the community is given by $d=d_1+d_2$ and the total number of edges is equal to $l_1+l_2+e$. An elementary calculation shows that the difference in the objective function between these two situations is  given by 
\begin{equation*}
\Delta Q = \frac{1}{L}  \left(e - \gamma \frac{d_1 d_2}{2L}\right)
\end{equation*}
with separate communities leading to a larger value of the objective function when $\Delta Q < 0$.

Since $e \leq d_1$, we have
\begin{equation*}
\Delta Q  \leq \frac{1}{L}  \left(d_1 - \gamma \frac{d_1 d_2}{2L}\right) = \frac{d_1}{L}  \left(1 - \gamma \frac{d_2}{2L}\right)
\end{equation*}
and so $\Delta Q <0$ when $d_2/(2L) > 1/\gamma$. Thus, if one can find a set of nodes in a community whose total node degrees  exceed $1/\gamma$ of the total node degrees in the network, then the value of the objective function increases when making this set of nodes a separate community. This imposes a non-trivial upper bound on community sizes as soon as $\gamma >1 $. In particular, a community of $n$ nodes may not contain a fraction of the total degree (or of the total number of edges) larger than $n/((n-1) \gamma)$.

We now show with an example that the use of a resolution parameter may disaggregate large and densely connected communities. Consider the network consisting of a clique of 16 nodes and of four cliques of 4 nodes each. There is one edge between the clique of 16 nodes and each of the cliques of 4 nodes. All pairs of cliques of 4 nodes are connected to each others with 2 edges (Figure 1). The partition of optimal modularity ($\gamma = 1$) consists of two communities of 16 nodes each, as shown on the left of Figure 1. This is a typical illustration of the resolution limit where modularity optimization fails to detect the four small cliques of 4 nodes. When $\gamma$  is increased to enable the detection of the smaller cliques, the larger clique of 16 nodes splits into 16 distinct communities of one node each (middle of Figure 1, $\gamma = 1.5$). As $\gamma$  is further increased, the second community finally splits into 4 cliques (right of Figure 1, $\gamma = 2$).

As this simple example clearly shows, when optimizing the objective function $Q_\gamma$ for $\gamma > 1$ one should be aware of the tendency of the resulting optimum to disagregate large and dense communities and be cautious when interpreting the partitions obtained.

\begin{figure}[!h]
    \centering
    \includegraphics[width=5cm,angle=0]{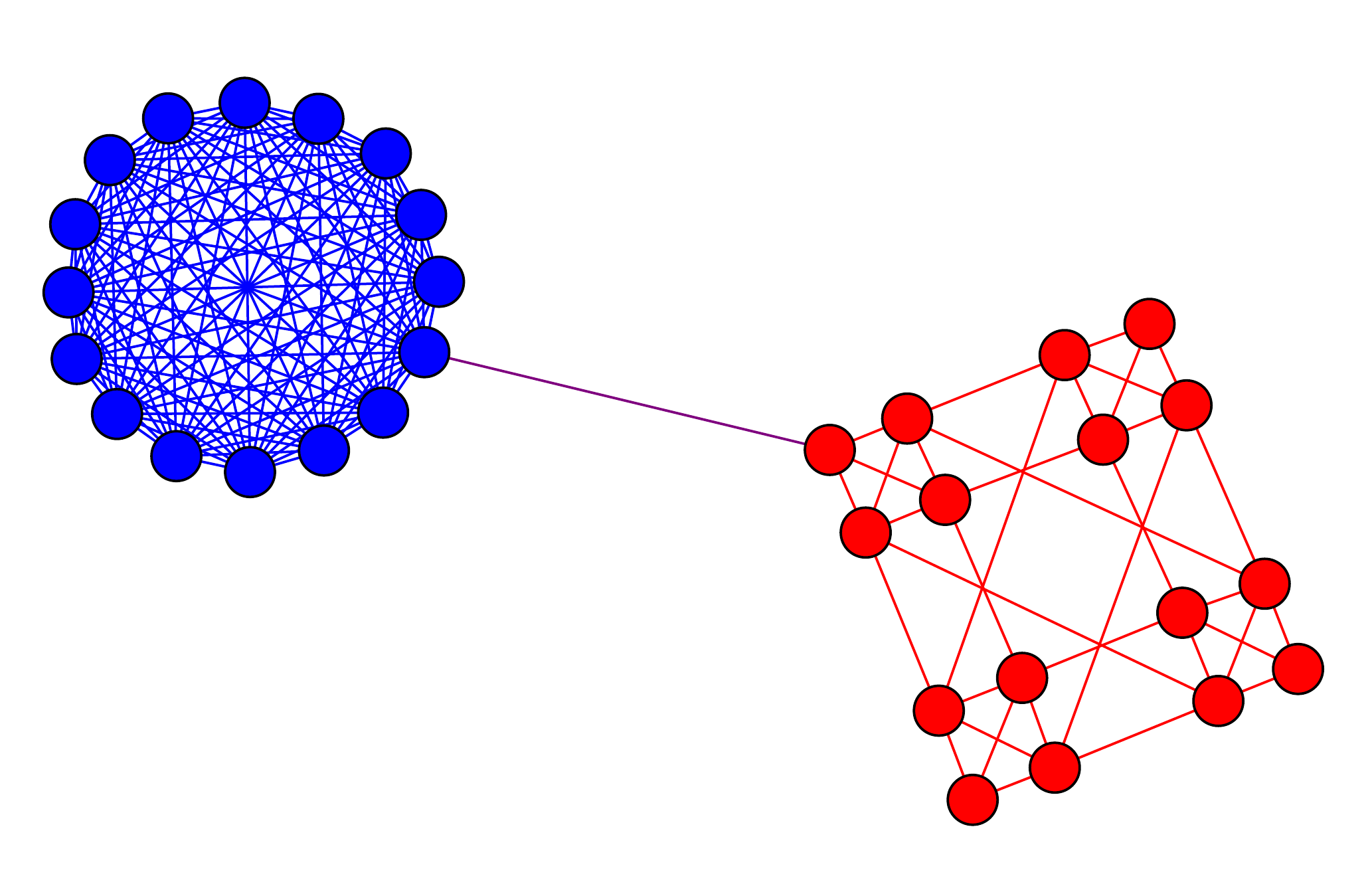}
     \includegraphics[width=5cm,angle=0]{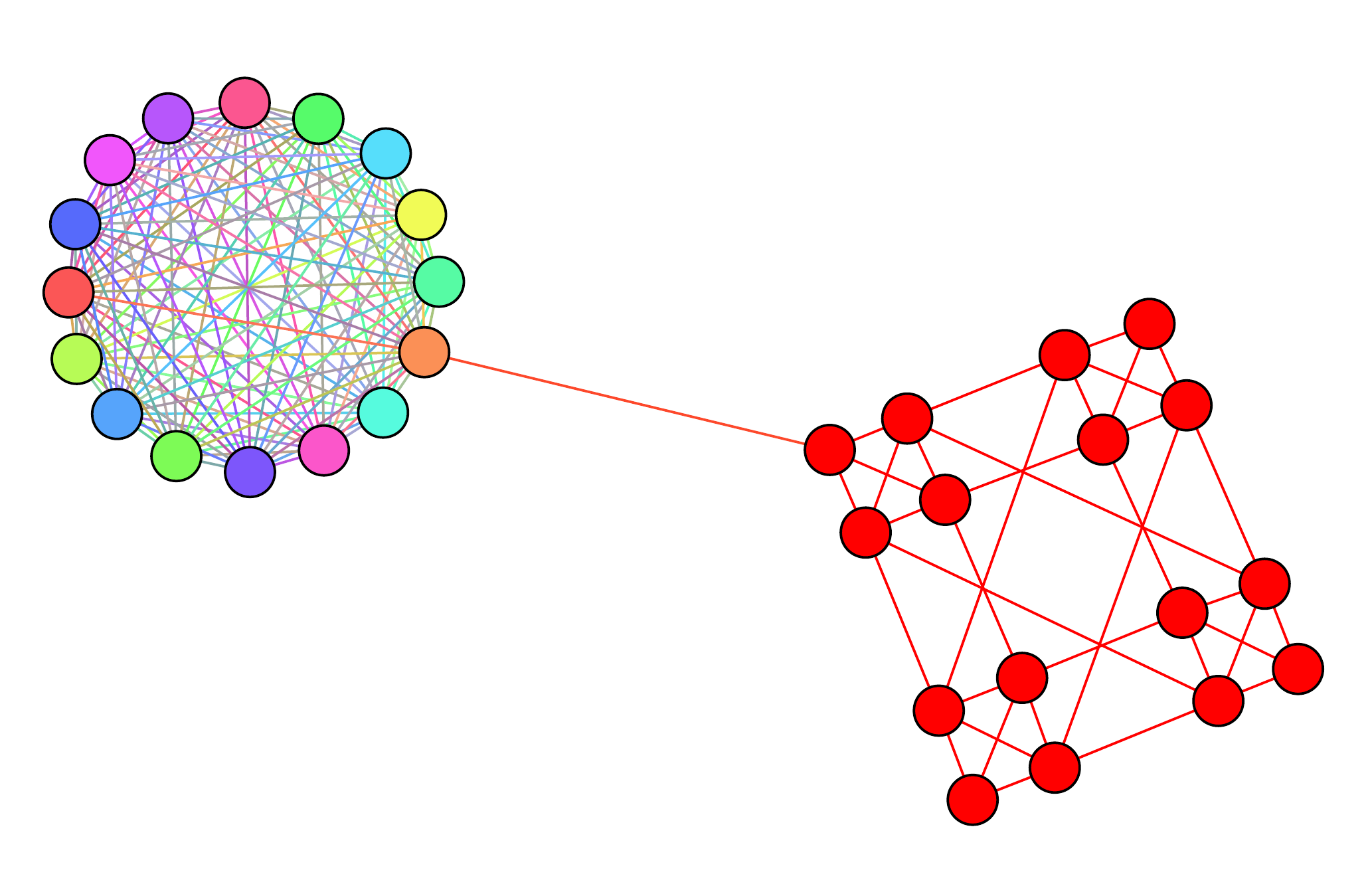}
 \includegraphics[width=5cm,angle=0]{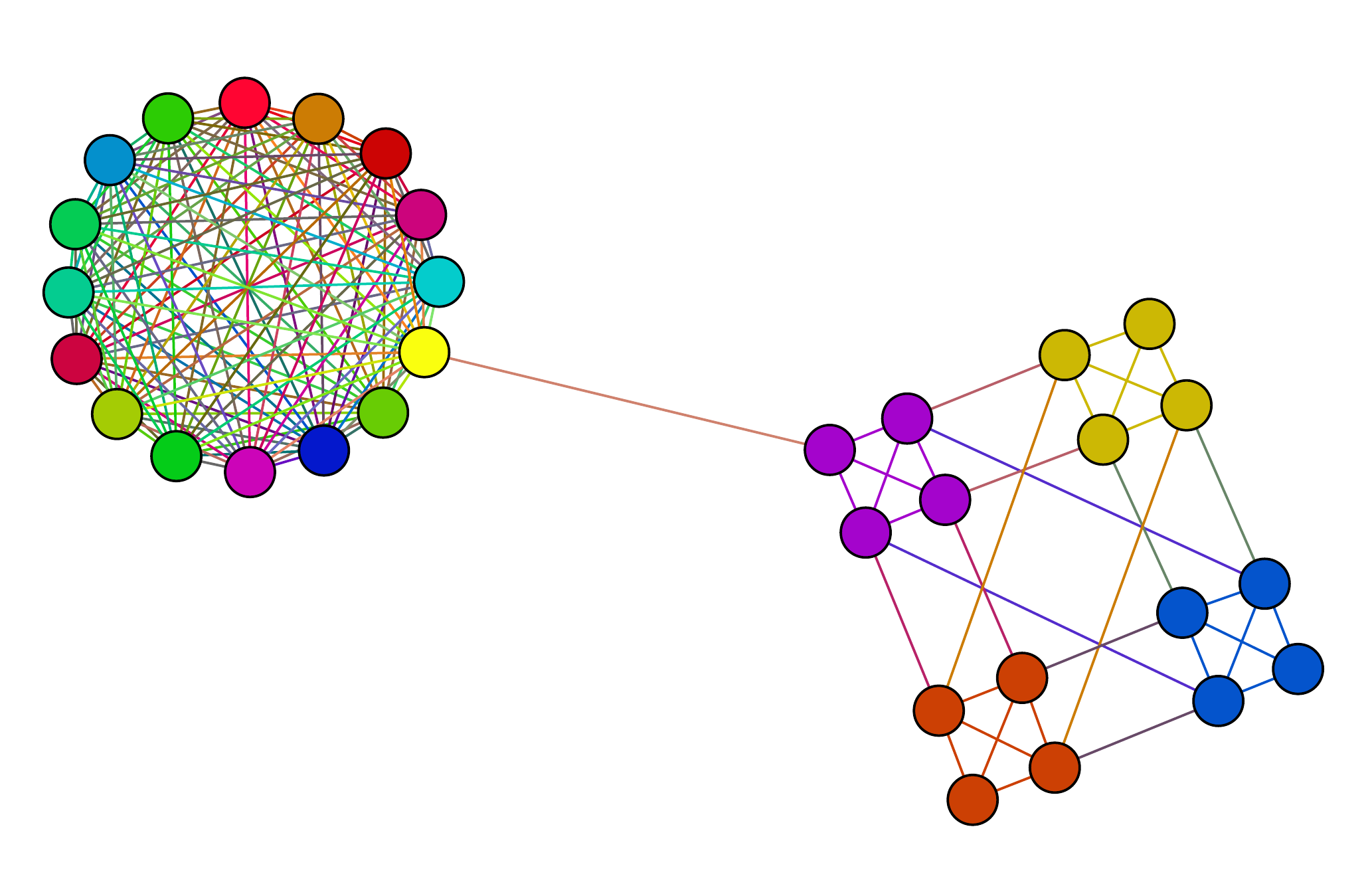}

   \caption{  Community partitioning of the same network for different values of the tunable resolution parameter $\gamma$  . On the left, the partition obtained for $\gamma$ = 1 consists of two communities of 16 nodes each ; this is a typical example of the resolution limit of modularity where modularity optimization fails to detect the four small cliques of 4 nodes. As the resolution parameter is increased to $\gamma$ = 1.5, the method still fails to detect the four small cliques but the nodes in the large clique now form sixteen distinct one-node communities. When $\gamma$ = 2 (right) the four small cliques are finally separated. }
\end{figure}

\begin{acknowledgments}
We wish to express our thanks to J.-C. Delvenne, R. Lambiotte, P. Mucha, V. Traag who have commented an earlier version of this note. 
\end{acknowledgments}

\bibliography{resol}

\end{document}